\documentclass[aps,prd,twocolumn,showpacs,amsmath]{revtex4}
\usepackage[dvips]{color,graphicx}
\usepackage{amsfonts}
\usepackage{amssymb}
\usepackage{latexsym}

\newcommand{\dalm}{\kern1pt\vbox{\hrule height 0.9pt\hbox{\vrule width
0.9pt\hskip 2.5pt\vbox{\vskip 5.5pt}\hskip 3pt\vrule width 0.3pt}\hrule height
0.3pt}\kern1pt}
\newcommand{\ma}[1]{\mbox{$\mathcal{#1}$}}

\newcommand{\lw}[1]{\smash{\lower2.ex\hbox{#1}}}

\begin{document}


\title{Characterization of the Lovelock gravity by Bianchi derivative}
\author{Naresh Dadhich\thanks{Electronic address:nkd@iucaa.ernet.in}}
\email{nkd@iucaa.ernet.in}
\affiliation{Inter-University Centre for Astronomy \& Astrophysics, Post Bag 4, Pune~411~007, India\\}
\date{\today}

\begin{abstract} 
We prove the theorem: The second order quasi-linear differential operator as a second rank divergence free tensor in the equation of motion for gravitation could always be derived from the trace of the Bianchi derivative of the fourth rank tensor, which is a homogeneous polynomial in curvatures. The existence of such a tensor for each term in the polynomial Lagrangian is a new characterization of the Lovelock gravity.  
\end{abstract}


\maketitle

The Bianchi differential identity involving the purely anti-symmetric derivative of a derivative ($D^2=0$) was famously interpreted by John Wheeler \cite{mtw} as the statement of the fact, {\it boundary of boundary is zero}. The familiar examples of it are curl of gradient and divergence of curl being zero. The former signifies a scalar while the latter a vector field. However its contraction in the two cases is vacuous and hence does not lead to a non-trivial statement$^{[1]}$\footnotetext[1]{For a vector field, vanishing of Bianchi derivative means the field is a curl of gauge potential;i.e. $F_{ik} = A_{i,k} - A_{k,i} = -F_{ki}$. This trivially implies $F^{ik}_{;ik} = 0$ and consequently we can write $F^{ik}_{;k} = J^i$ with $J^i_i = 0$. This is indeed the equation of motion for the gauge field with $J^i$ being identified as the current vector.}. When we go beyond vector to a tensor field, it becomes interesting. \\

Gravity is the universal force which means it links to all particles unmindful  of their mass being non-zero or zero. It is its linkage to massless particles which leads to the profound realization that it could only be described by curvature of spacetime \cite{nd1,nd2}. This means the dynamics of gravity resides in spacetime curvature which must fully and entirely determine it. That is the gravitational dynamics must follow from the curvature which is described by the fourth rank Riemann curvature tensor (it is defined as $A^i_{;lk} - A^i_{;kl} = R^i{}_{mlk}A^m$, a generalized ``curl''). It involves second and first derivatives of the spacetime metric, $g_{ab}$. The Bianchi identity is given by the anti-symmetric derivative of $R^i{}_{mlk}$,
\begin{equation}
R^i{}_{m[lk;n]} = 0 . 
\end{equation}
If the gravitational dynamics has to follow from the curvature, it has to follow from this identity which is the only available geometric relation. The only thing we can do to it is to contract on the available indices which does lead, unlike for scalar and vector, to a non-vacuous relation, 
\begin{equation}
G^{a}{}_{b;a} = 0, ~~~~ G_{ab} = R_{ab} - {1\over2}Rg_{ab} 
\end{equation}
where $R_{ab}$ is the Ricci tensor, the contraction of Riemann, while $R$ is the trace of Ricci. Now the trace (contraction) of the Bianchi identity yields a non-trivial differential identity from which we can deduce the following relation 
\begin{equation}
G_{ab} = \kappa T_{ab} - \Lambda g_{ab}, ~~~ T^{a}{}_{b;a} = 0  
\end{equation}  \\
where $T_{ab}$ is the second rank symmetric tensor with vanishing divergence and $\kappa$ and $\Lambda$ are constants. The left hand side of the equation is a second order differential operator on the metric $g_{ab}$. For this equation to describe dynamics of gravity, the tensor $T_{ab}$ should describe the source/charge for gravity which should also be universal. It should be shared by all particles and hence $T_{ab}$ should represent energy momentum distribution. Thus we obtain the Einstein equation for gravitation which entirely follows from the spacetime curvature. We have however two constants of which one $\kappa$ is to be determined by experimentally measuring the strength of the force and is identified with Newton's constant, $\kappa = -8 \pi G/c^2$. Why is there new constant $\Lambda$ which though arises in the equation as naturally as the energy momentum tensor, $T_{ab}$? It is perhaps because of the absence of fixed spacetime background which exists for the rest of physics and the new constant may be a signature of this fact. It is the universal character of gravity which makes spacetime itself dynamic. The force free state would however be characterized by homogeneity and isotropy of space and homogeneity of time which will in general be described by spacetime of constant curvature and not necessarily of zero curvature. The new constant $\Lambda$ is the measure of the constant curvature of spacetime and it identifies the most general spacetime for force free state. It may in some deep and fundamental sense be related to the basic structure of spacetime. \\

We also know that the complete contraction of Riemann gives the scalar curvature, $R$, the Einstein-Hilbert Lagrangian which on variation leads to the divergence free Einstein tensor, $G_{ab}$, and subsequently the Einstein equation. We thus have the two distinct but equivalent derivations for the gravitational dynamics. The former is simply driven by the geometry while the latter is in the spirit of every dynamical equation following from an action. It is always possible to write an action constructed from Riemann curvature for higher derivative gravity and derive the corresponding equation of motion. Similarly, is it possible to derive an analogue of $G_{ab}$, a divergence free differential operator from the Bianchi derivative of the higher order curvature polynomial? This is the question we wish to address and show that the answer is in affirmative. It would give yet another characterization of the Lovelock gravity. \\
  
We believe that gravitational dynamics to follow from the spacetime curvature should be a general principle which should be true in general for higher order theories as well. So comes the question of going beyond the linear order in Riemann. Let us consider the quadratic tensor,
\begin{equation}  
\ma R_{abcd} = R_{abmn}R_{cd}{}^{mn} + \alpha R_[{a}{}^{m}R_{b]mcd} + \beta R R_{abcd} \label{gb1}
\end{equation}
where $\alpha, \beta$ are constants. We consider the Bianchi derivative, $\ma R_{ab[cd;e]}$ which on contraction gives
\begin{equation}  
g^{ac} g^{bd} \ma R_{ab[cd;e]} = (-2\ma R_e{}^c + \ma R \delta_e{}^c)_{;c} \label{bd1}
\end{equation}
where $\ma R_{ac} = g^{bd} \ma R_{abcd}$ and $ \ma R = g^{ab} \ma R_{ab}$. It turns out that for $\alpha=4$ and $\beta=1$, we obtain 
\begin{eqnarray}  
\ma R^{cd}{}{}_{[cd;e]}&=& (-2\ma R_e{}^c + \ma R \delta_e{}^c)_{;c} \nonumber \\
&=& (-H_e{}^c + {1\over2} L_{GB} \delta_e{}^c)_{;c}. 
\end{eqnarray}
That is
\begin{equation}
\ma R^{cd}{}{}_{[cd;e]} - {1\over2} (L_{GB} \delta_e{}^c)_{;c} = -H_e{}^c_{;c} = 0.\label{bd2}
\end{equation}
The tensor $H_{ab}$ is divergence free, $H_a{}^b_{;b} = 0$, and is given by 
\begin{eqnarray}
H_{ab}& =& 2(RR_{ab} - 2R_{a}{}^{m}R_{bm} - 2R^{mn}R_{ambn}\nonumber \\  
&+& R_{a}{}^{mnl}R_{bmnl}) - \frac{1}{2} L_{GB} g_{ab}.
\end{eqnarray}
It results from the variation of the well-known Gauss-Bonnet Lagrangian $L_{GB} = R_{abcd}^2 - 4R_{ab}^2 + R^2$ where we have written $R_{abcd}^2 = R_{abcd}R^{abcd}$. That is, we can write 
\begin{equation}  
H_{ab} = 2\ma R_{ab} - \frac{1}{2}\ma R g_{ab} \label{hab}
\end{equation}
where $\ma R = L_{GB}$.\\ 

Though $H_{ab}$ can be written in terms of $\ma R_{abcd}$ but it doesn't follow directly from it as $G_{ab}$ does from $R_{abcd}$. We note that Bianchi derivative vanishes only for the Riemann curvature signifying the fact it can be written as a generalized curl of a vector. No other tensor will have vanishing Bianchi derivative. However to find the  analogue of $G_{ab}$, we donot require vanishing of Bianchi derivative of the quadratic tensor $\ma R_{abcd}$ but instead vanishing of its trace would suffice. We see that even the trace of Bianchi derivative does not vanish but is equal to ${1\over2} (L_{GB} \delta_e{}^c)_{;c}$. It suggests that the curvature polynomial should also include term involving its 
own trace which would make no contribution in the linear case. Before we do that, let us write $\ma R_{abcd}$ and $H_{ab}$ for a general order $n$ in the Lovelock polynomial ~\cite{lov}. In the context of higher derivative gravity theories, a unified scheme of writing Lagrangian is given ~\cite{paddy} as an invariant product, $Q^{abcd}R_{abcd}$ where $Q^{abcd}$ has the same symmetry properties as the Riemann tensor. It is constructed from metric and Riemann curvature and has vanishing divergence, $Q^{abcd}{}{}{}{}_{;c}=0$. For the Lovelock Lagrangian, it is required to be a homogeneous function of the Riemann curvature. Since $\ma R_{abcd}$ is also a homogeneous polynomial in Riemann curvature for a given order, it is hence possible to write
\begin{equation}
\ma R_{abcd} = Q_{ab}{}^{mn}R_{mncd} \label{new}
\end{equation}
where $Q_{abcd}$ is in general given for $n$-th order by ~\cite{paddy},
\begin{equation}
Q^{ab}{}_{cd} = \delta^{aba_1b_1...a_nb_n}_{cdc_1d_1...c_nd_n}R_{a_1b_1}{}^{c_1d_1}...R_{a_nb_n}{}^{c_nd_n}. \label{new1}
\end{equation}
Let us verify this for the quadratic Gauss-Bonnet case where $Q_{abcd}$ could explicitly be written as ~\cite{paddy},
\begin{equation}
Q_{abcd} = R_{abcd} - 2R_{a[c}g_{d]b} + 2R_{b[c}g_{d]a} + Rg_{a[c}g_{d]b}. 
\end{equation}
It is easy to see that when it is substituted in Eqn. (\ref{new}), we obtain 
the Gauss-Bonnet $\ma R_{abcd}$ as given in Eqn. (\ref{gb1}) with $\alpha=4, \beta=1$. We could thus write $\ma R_{abcd}$ by using Eqs (\ref{new1}) and (\ref{new}) for any term in the Lovelock polynomial and the corresponding $H_{ab}$ for the $n$-th polynomial will be given by
\begin{equation}  
H_{ab} = n\ma R_{ab} - \frac{1}{2} \ma R g_{ab} \label{genH} 
\end{equation} 
which will for the linear case $n=1$ be the Einstein tensor, $G_{ab}$. In this case, the trace of the Bianchi derivative vanishes because the Riemann tensor satisfies the Bianchi identity. However $H_{ab}$ as defined above is the analogue of the Einstein tensor for $n$-th order polynomial and hence the generalized Einstein tensor is divergence free. As trace of $R_{abcd}$ is the Einstein-Hilbert Lagrangian similarly trace of $\ma R_{abcd}$ gives the Gauss-Bonnet and higher order Lovelock Lagrangian. Note that Eqn. (\ref{bd1}) holds good in 
general for any order $n$ of the polynomial $\ma R_{abcd}$. We can thus write in general for Eqn. (\ref{bd2}),
\begin{equation}
\ma R^{cd}{}{}_{[cd;e]} - \frac{n-1}{n}(\ma R \delta_e{}^c)_{;c} = -\frac{2}{n}{}{} H_e{}^c_{;c} = 0 \label{bd3}
\end{equation}
where $\ma R$ is the corresponding $n$-th order Lagrangian. For $n=1$, the second term on the left vanishes indicating the Bianchi identity and $H_{ab}=G_{ab}$. For $n>1$, the trace of Bianchi derivative does not vanish unless the trace $\ma R$ is also included as indicated by the second term which then leads to the required divergence free tensor, $H_{ab}$. Apart from the quadratic Gauss-Bonnet case, we have also verified the above relation for $n=3$ cubic polynomial. It is interesting that the order of polynomial appears in the definition of the corresponding curvature polynomial tensor yielding the second rank symmetric divergence free tensor. \\  

We now turn to redefining the curvature polynomial which also includes its trace such that trace of the Bianchi derivative vanishes directly giving the required differential operator in terms of divergence free $H_{ab}$. We thus write 
\begin{equation}
\ma F_{abcd} = \ma R_{abcd} - \frac{n-1}{n(d-1)(d-2)} \ma R (g_{ac}g_{bd} - g_{ad}g_{bc}) \label{F}
\end{equation}
where $n$ being the order of the polynomial and $d$ the dimension of spacetime under consideration. Then 
\begin{equation}
 -\frac{n}{2} \ma F^{cd}{}{}_{[cd;e]} = H_e{}^c_{;c} = 0  
\end{equation} 
which  verifies Eqn. (\ref{genH}). In terms of $\ma F_{abcd}$, $H_{ab}$ is given by
\begin{equation}
n(\ma F_{ab} -\frac{1}{2} \ma F g_{ab}) = H_{ab} 
\end{equation}
where $\ma F_{ab} = g^{cd} \ma F_{acbd}$ and $\ma F = g^{ab} \ma F_{ab}$. It is interesting to note that  $\ma F_{ab}$ is the analogue of the Ricci, $R_{ab}$ and $H_{ab}$ of the Einstein, $G_{ab}$ with the order $n$ of the polynomial $\ma R_{abcd}$ which is the analogue of the Riemann, $R_{abcd}$. Thus $H_{ab}$ is obtained  from $\ma F_{abcd}$ in the same manner as $G_{ab}$ is from $R_{abcd}$, and for $n=1$, $\ma F_{abcd} = \ma R_{abcd} = R_{abcd}$. All this however happens only for the specific Lovelock coefficients in the polynomial and thereby determining the Lovelock polynomial. \\ 

Note that $H_{ab}$ defines the general conserved (Einstein) tensor for any order $n$ of the homogeneous curvature polynomial and let us take its trace which would give  
\begin{equation}
\ma F = \frac{d-2n}{n(d-2)} \ma R. 
\end{equation}
For the usual Einstein gravity, $n=1$, it is $\ma F = \ma R$ with $d>2$ and the Einstein tensor vanishes for $d=2$. Thus we have the general result that $d>2n$ for $n$-th order polynomial and $H_{ab}$ vanishes for $d=2n$. It is the variation of $\ma R$ (and so also $\ma F$) gives $H_{ab}$ which vanishes for $d=2n$ but not $\ma R$. If we take $\ma F$ which is proportional to the Lovelock Lagrangian $\ma R$, then the Lagrangian itself vanishes for $d=2n$. It is remarkable that the existence of an analogue of the Einstein tensor in general - a conserved tensor which is non-zero only for $d>2n$, uniquely determines the Lovelock polynomial. \\

By redefining the curvature polynomial we have been able to derive the divergence free differential operator for the gravitational dynamics. This is what we had set out to do and it could be stated as follows: \\

{\bf Theorem:} {\it The second order quasi-linear differential operator as a second rank divergence free tensor in the  equation of motion for gravitation could always be derived from the trace of the Bianchi derivative of the fourth rank tensor, $\ma F_{abcd}$, which is a homogeneous polynomial in curvatures. The trace of the curvature polynomial is proportional to the corresponding term in the Lovelock action and corresponding to each term in the Lovelock Lagrangian, there exists a fourth rank tensor which is a new characterization of the Lovelock Lagrangian. } \\

It is the requirement of quasi-linearity (linear in second derivative) of the equation of motion which singled out the Lovelock polynomial with specific coefficients. In our case it is replaced by the requirement that the fourth rank tensor which is homogeneous in curvatures yields a divergence free second rank tensor through the trace of its Bianchi derivative. This automatically ensures the quasi-linearity of the equation. For non Lovelock action, there would not exist a fourth rank tensor with this property.  \\

In a sense, the Riemann curvature could be looked upon as a {\it Bianchi potential} giving the Einstein tensor for the Einstein gravity while $\ma F_{abcd}$ is the {\it Bianchi potential} giving $H_{ab}$ for the Lovelock dynamics. Thus each term in the Lovelock polynomial has a potential tensor $\ma F_{abcd}$ given by Eqs (\ref{new}), (\ref{new1}) and (\ref{F}). It is remarkable that there exists {\it Bianchi potential} for each term in the Lovelock Lagrangian. Non existence of potential characterizes all other actions like $f(R)$ gravity. The existence of {\it Bianchi potential} thus becomes a distinguishing feature for the Lovelock action. In this context it may be noted that the requirement that both Palitini and metric action give the same equation of motion also picks up the Lovelock action ~\cite{jab}. We have thus three distinct properties (namely quasi-linearity of equation of motion, equivalence of Palitini and metric formulation and existence of Bianchi potential) which characterize the Lovelock action. I believe that there should exist a thread knitting them and it would be interesting to probe that. \\  

However, the relevance of the order in the Lagrangian for the gravitational dynamics depends upon the spacetime dimension. For instance, for $d<5$, the quadratic Gauss-Bonnet term makes no contribution to the equation of motion and similarly the cubic term becomes relevant only for $d>6$. We would however like to emphasize that for $d>4$, $H_{ab}$ must be included along with $G_{ab}$ for the classical dynamics of gravitation. In the ultra violet limt of the theory signifying super strong gravitational field, it is pertinent to include higher order curvature effects. If we continue to have a well defined evolution of the field, the equation must be quasi-linear. This will inevitably and uniquely lead to Lovelock polynomial and higher dimension.  Further one loop correction in string theory generates the Gauss-Bonnet term \cite{st}. That is, strong field limit of classical gravity and one loop quantum correction seem to share the same Gauss-Bonnet ground. It could therefore be envisioned that the classical limit to quantum gravity is perhaps via the Lovelock gravity and the relevant order in the polynomial (loop correction) being given by the spacetime dimension under consideration. The Gauss-Bonnet and higher orders may therefore represent a intermediary state between the classical Einstein gravity and quantum gravity \cite{dad1}.\\

Gravity is an inherently self interactive force and the self interaction could only be evaluated by successive iterations. The Einstein gravity is self interactive but it contains only the first iteration through the square of first derivative in Riemann curvature. The question is how do we stop at the first iteration? The second iteration would ask for a quadratic polynomial in Riemann curvature which should give the corresponding term in the equation of motion. Thus the quadratic tensor, $\ma R_{abcd}$ as given in Eqn. (\ref{gb1}) with specific coefficients, will alone meet the requirement (quasi-linearity of the differential operator) for inclusion of the second iteration. However its effect in the equation of motion can be felt only for dimension, $d>4$, and hence we have to go to higher dimension for physical realization of the second iteration of self interaction \cite{dad1,dad2}. It is remarkable that even classical dynamics of gravity asks for dimension $>4$. As two and three dimensions were not big enough for free propagation of gravity, similarly four dimension is not big enough to fully accommodate self interaction dynamics of gravity. Then the most pertinent question is where does this chain end? Let us envision the AdS/CFT-like scenario where the $3$-brane ($(3+1)$-spacetime) forms the boundary enclosing the higher dimensional bulk spacetime. If matter fields remain confined to $3$-brane, the bulk would then be free of matter and hence it would be maximally symmetric (homogeneous and isotropic in space and homogeneous in time). It is then a spacetime of constant curvature, dS/AdS, with vanishing Weyl curvature. There is no free gravity to propagate any further and hence the chain stops at the second level at least in this particular construction. Whether this is a generic setting or not is however an open question? It may be noted here that gravitational equation with inclusion of both $G_{ab}$ and $H_{ab}$ for empty space in higher dimension admits dS/AdS as solution \cite{dad1}. \\

In what dimension should matter live is to be entirely determined by the matter dynamics. If the matter fields are gauge vector fields described by $2$-form, the conformal invariance (universal scale change, $g_{ab}\to f^2 g_{ab}$) of their dynamics will dictate that they live in four dimension. This general principle is always obeyed by the matter field dynamics unless a scale in terms of mass etc. is introduced by spontaneous symmetry breaking. The symmetry breaking is indicative of theory being incomplete and it is hoped that the complete theory would restore the conformal invariance. It is therefore reasonable to take that matter remains confined to $3$-brane. This view is also supported by the string theory where open strings have their endpoints on the brane indicating residence of matter there. \\

The most fundamental question is how do we know there exist higher dimensions and if so why do we not see them? The existence of dimension can only be probed by a physical interaction. All our probes are matter field forces like electromagnetic, which remain, as argued above, confined to $3$-brane and hence they cannot fathom higher dimension. Since gravity is universal and hence it cannot be confined entirely to the brane and can propagate in higher dimension. Above we have argued quite convincingly that there is a strong case for higher dimension for physical realization of the second iteration of self interaction. It is a purely classical motivation (we have also elsewhere \cite{dad1,dad2} given a couple of more classical arguments) while higher dimension is a natural arena for string theory. The only way higher dimension could thus be probed is by a very high energy pure gravitational experiment. This is what we have not yet been able to do and hence the question remains open. \\   
         
We had set out to establish the general principle that gravitational dynamics resides in spacetime curvature and hence it should always and entirely be driven by spacetime geometry. This we have shown by deriving the quasi-linear differential operator for the equation of motion for the Lovelock gravity. In particular, we have found a new geometric characterization of the Lovelock gravity in existence of {\it Bianchi potential} for each term in the polynomial. This is indeed an interesting general property.  \\

{\it Acknowledgment:} I wish to thank T Padmanabhan and Dawood Kothawala for useful discussion and the latter also for Eqn. (\ref{genH}). 


\end{document}